\documentclass[onecolumn,showpacs,preprintnumbers]{revtex4}
\usepackage{graphicx}
\begin{document}

\title{Generalization of the DMPK Equation beyond Quasi One
Dimension}

\author{K. A. Muttalib}%
\author{V\'{\i}ctor A. Gopar}
\altaffiliation{Present address: 23, rue du Loess, F-67037 Strasbourg
Cedex (France)} %
\affiliation{Department of Physics, University of
Florida, P.O. Box 118440, Gainesville, FL 32611-8440.}

\begin{abstract}
Electronic transport properties in a disordered quantum wire are
very well described by the Dorokhov-Mello-Pereyra-Kumar (DMPK)
equation, which describes the evolution of the transmission
eigenvalues as a function of the length of a multichannel
conductor. However, the DMPK equation is restricted
to quasi one dimensional systems only. We derive a generalized
DMPK equation for higher dimensions, containing dependence on the
dimensionality through the properties of the transmission
eigenvectors, by making certain statistical assumptions about the
transfer matrix. An earlier phenomenological generalization is
obtained as a special case.
\end{abstract}


\maketitle

The Dorokhov-Mello-Pereyra-Kumar (DMPK) equation
\cite{dorokhov,mpk} has been enormously successful in describing
the electron transport properties of a quasi one dimensional
disordered conductor \cite{beenakker}. The equation describes the
evolution of the joint probability distribution of the
transmission eigenvalues with increasing length of the system, and
has been shown to be equivalent \cite{frahm} to the description in
terms of a non-linear sigma model \cite{efetov} obtained from the
microscopic tight binding Anderson Hamiltonian. The advantage of
the DMPK approach over the non-linear sigma model is that one can
consider the full distribution of transport quantities rather than
the mean and the variance alone. Recent analytical as well as
numerical results show that the distribution of conductances
\cite{mu-wo} has many surprises, including very sharp features at
(dimensionless) conductance $g=1$ which could not be anticipated
from studies of the moments of the distribution, and which should
have important consequences for the Anderson transition. One major
disadvantage of the DMPK equation, however, is that it does not
contain information about the spatial structure of the sample in
directions perpendicular to the direction of the current flow,
limiting its applicability to quasi one dimension (Q1D) only.
Since at present there is no other analytic approach available to
study the full distribution of transport properties, it is clear
that a generalization of the DMPK equation valid in higher
dimensions is of fundamental importance. A phenomenological
generalization, with an ad hoc constraint to conserve probability,
was recently proposed \cite{mu-kl} which seems to agree with
numerical results \cite{markos} in systems beyond Q1D in some
restricted regimes. In the present paper we derive a further
generalization that contains dependence on the dimensionality
through the properties of the transmission eigenvectors, and
contains the earlier model as a special case. No additional
constraint is needed to conserve probability in the present
approach. Moreover, the approach reproduces the expression for
Lyapunov exponents in higher dimensions obtained in
[\onlinecite{chalker}]. Known properties of these exponents
provide useful constraints on the phenomenological parameters in
the current model.

\begin{figure}
\includegraphics[angle=0,width=0.39\textheight]{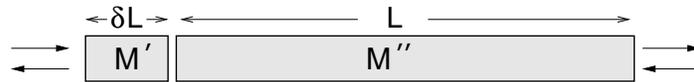}
\caption{\label{fig1} A small wire with length $\delta L$, ``the
building block'', is attached to a long one with length $L$
($\delta L \ll L$). The total transfer matrix $M$ is given by
$M=M''M'$ .}
\end{figure}

In the transfer matrix approach, a conductor of length $L$ is
placed between two perfect leads of finite width. The scattering
states at the Fermi energy define $N$ channels. The $2N \times 2N$
transfer matrix $M$ relates the flux amplitudes on the right of
the system to that on the left \cite{muttalib}. Flux conservation
and time reversal symmetry (we consider the case of unbroken time
reversal symmetry only) restricts the number of independent
parameters of $M$ to $N(2N+1)$ and can be written in general as
\cite{mpk}
\begin{equation}
M=\left(\matrix{ u & 0  \cr 0 & {u}^* \cr }\right)
\left(\matrix{ \sqrt{1+\lambda} & \sqrt{\lambda}   \cr
\sqrt{\lambda}   & \sqrt{1+\lambda} \cr }\right)\left(\matrix{ v &
0  \cr 0 & {v}^* \cr }\right)\equiv U \Gamma V,
\end{equation}
where $u,v$ are $N \times N$ unitary matrices, and $\lambda$ is a
diagonal matrix, with positive elements $\lambda_i, i=1,2, ...N$.
An ensemble of random conductors of length $L$, all with the same
macroscopic disorder characterized by the same mean
free path $l$ but different microscopic realizations of the
randomness, is described by an ensemble of random transfer
matrices $M$, whose differential probability depends
parametrically on $L$ and can be written as
$dp_L(M)=p_L(M)d\mu(M)=p_L(\Gamma,U,V)d\mu(\Gamma)d\mu(U)d\mu(V)$.
Here $d\mu(M)$ is the invariant Haar measure of the group, given
in terms of the parameters in (1) by
$d\mu(M)=J(\lambda)\left[\prod_{i}^{N}d\lambda_i \right] d\mu(u)
d\mu(v)$, where $J(\lambda)=\prod_{i<j}|\lambda_i-\lambda_j|^{\beta}$
with the `symmetry exponent' $\beta=1$ in our case, and
$d\mu(u)$ and $d\mu(v)$ are the invariant measures of the unitary
group $U(N)$.

We now take two conductors, with lengths $L$ and $\delta L$, to
make a conductor of length $L+\delta L$ (Fig. 1). Denoting the
corresponding transfer matrices by $M''$, $M'$ and $M=M''M'$ with
probability densities $p_L(M''=MM'^{-1})$, $p_{\delta L}(M')$ and
$p_{L+\delta L}(M)$, respectively, we have the relation
\cite{mpk}:
\begin{equation}
p_{L+\delta L}(M)=\int p_L(MM'^{-1})p_{\delta L}(M')d\mu (M').
\end{equation}

The restriction of the DMPK equation to Q1D arises from the
``isotropy'' approximation, that the distribution $p_L(M)$ is
independent of the matrices $u$ and $v$. Several attempts have
been made in order to relax the isotropy approximation
\cite{mello,chalker,mu-kl}. We will avoid writing a set of coupled
evolution equations for $\lambda$, $u$ and $v$ by considering the
marginal distribution
\begin{equation}
\bar{p}_{L}(\lambda)= \int p_{L}(\Gamma,U,V)d\mu(U)d\mu(V).
\end{equation}
We first show that the $U$ integral can be done exactly and then
we make statistical assumptions about certain products of the
matrix elements of $V$.

Consider the combination $H=MM^{\dag}=U\Gamma^2U^{\dag}$. At
length $L$, we get $H''= M''M^{''\dag} = U\Gamma
VM^{'-1}(M^{'-1})^{\dag}V^{\dag}\Gamma U^{\dag}\equiv
U''\Gamma''^2U''^{\dag}.$ It then follows that we can write $U''=
U\cdot U'''(\Gamma,V,M')$. Similarly, by considering the
combination $Q = M^{\dag}M = V^{\dag}\Gamma^2 V$, we get $Q'' =
M^{''\dag}M'' = (M^{'-1})^{\dag}V^{\dag}\Gamma^2 VM^{'-1} \equiv
V''^{\dag}\Gamma''^2V'',$ which implies that we can write
$\Gamma'' = \Gamma''(\Gamma,V,M')$ and $V'' = V''(\Gamma,V,M')$.
Eq. (2) can then be rewritten as
\begin{eqnarray}
p_{L+\delta L}(U,\Gamma,V)&=&\int
p_L(U\cdot U'''(\Gamma,V,M'),\Gamma''(\Gamma,V,M'),V''(\Gamma,V,M'))\cr
&\times & p_{\delta L}(M')d\mu (U')d\mu(\Gamma')d\mu(V').
\end{eqnarray}
We now integrate both sides of Eq. (4) over the invariant measure
$d\mu(U)$. The left hand side defines a marginal distribution
$q_{L+\delta L}(\Gamma,V)$. On the right side, since the measure
is invariant, $d\mu(U)=d\mu(U \cdot U''')$ for $U'''$ fixed.
Therefore $d\mu(U)=d\mu(U'')$, and integrating over $d\mu(U'')$
gives the marginal distribution $q_L(\Gamma'',V'')$ with the
following convolution equation:
\begin{equation}
q_{L+\delta L}(\lambda,v)=\int q_L(\lambda'',v'') p_{\delta
L}(\lambda',u',v')d\mu (M'),
\end{equation}
where we have used Eq. (1) to introduce the matrix elements
$\lambda, v, \lambda',v'$, etc.

Writing $\lambda''=\lambda+\delta\lambda$ and $v''=v+\delta v$,
Eq. (5) can be rewritten as
\begin{equation}
q_{L+\delta L}(\lambda, v)= \left<
q_{L}(\lambda+\delta\lambda(\lambda,v), v+\delta
v(\lambda,v))\right>_{\delta L} ,
\end{equation}
where $\left<...\right>_{\delta L}$ denotes an average over the
ensemble of $M'$. In order to obtain $\delta \lambda$ and $\delta
v$ within a perturbation theory, we search for a matrix
constructed from $M$ whose eigenvalues and eigenvectors are given
by $\lambda$ and columns of $v$, respectively. Consider the matrix
$Q=M^{\dagger}M=V^{\dag}\Gamma^2V$. Flux conservation implies
$Q^{-1}=\Sigma_zQ\Sigma_z$, where $\Sigma_z$ is the $2N\times 2N$
generalization of the Pauli matrix $\sigma_z$. It then follows
that the matrix $X=[Q+Q^{-1}-2I]/4$, where $I$ is the identity
matrix, is block diagonal. It has been shown that $V$ diagonalizes
$X$ \cite{muttalib}, leading to $N$ doubly degenerate eigenvalues
$\lambda$. We can therefore obtain $\delta \lambda$ and $\delta v$
by considering the change $\delta X = X''-X$ due to change in $M$
arising from $M'$. Writing $\delta Q=Q''-Q$, it is easy to see
that $\delta X$ is also block diagonal, with $\delta
X_{11}=\frac{1}{2}\delta Q_{11}$ and $\delta
X_{22}=\frac{1}{2}\delta Q_{11}^*$. Since both $X$ and $\delta X$
are block diagonal, the perturbation can be treated as acting
separately on the two sub-blocks of $X$, and one can use ordinary,
as opposed to degenerate, perturbation theory to obtain $\delta
\lambda$ and $\delta v$ by considering one sub-block only.

Denoting the perturbation by $\tilde{w}=v(\delta
X_{11})v^{\dagger}=\frac{1}{2}v(\delta Q_{11})v^{\dagger}$ we get
$$
\tilde{w}=-\lambda+
vu'\left[\lambda'+\sqrt{1+\lambda'}v'v^{\dagger}\lambda v
v'^{\dagger} \sqrt{1+\lambda'}
+\sqrt{\lambda'}v'^*v^{*\dagger}\lambda v^*
v'^{*\dagger}\sqrt{\lambda'} \right]u'^{\dagger}v^{\dagger}
$$
\begin{equation}
-vu'[\sqrt{1+\lambda'}v'v^{\dagger}\sqrt{\lambda(1+\lambda)}
v^*v'^{*\dagger}\sqrt{\lambda'}+\sqrt{\lambda'}v'^*v^{*\dagger}
\sqrt{\lambda(1+\lambda)}vv'^{\dagger}\sqrt{(1+\lambda'}]
u'^{\dagger}v^{\dagger}.
\end{equation}
We expect $\lambda'\propto\delta L/l << 1$. Since $\tilde{w}$
contains terms proportional to $\sqrt{\lambda'}$, we need to
consider corrections to both $\lambda$ and $v$ up to second order
in $\tilde{w}$ in order to keep terms up to $O(\lambda')$.
Standard perturbation theory \cite{landau} gives $\delta
\lambda_a=\delta \lambda^{(1)}_a+\delta \lambda^{(2)}_a$ and
$\delta v^{\dag }_{an}= \delta v^{\dag (1)}_{an}+\delta v^{\dag
(2)}_{an}$ with
\begin{equation}
\delta \lambda^{(1)}_a= \tilde{w}_{aa}; \;\;\;\delta
\lambda^{(2)}_a= \sum_{b(\ne
a)}\frac{\tilde{w}_{ab}\tilde{w}_{ba}}{\lambda_a-\lambda_b},
\end{equation}
\begin{equation}
\delta v^{\dag (1)}_{an}= \sum_{m(\ne
n)}\frac{\tilde{w}_{mn}}{\lambda_n-\lambda_m}v^{\dag}_{am}
\end{equation}
and similarly for $\delta v^{\dag (2)}_{an}$. Note that $v+\delta
v$ has to remain unitary, which imposes an additional constraint.

The averages over the ``building block'' $M'$ appearing in Eq. (6)
involve averages over combinations of $\lambda'$, $u'$ and $v'$
which appear in $\tilde{w}$. Note that the building block is
highly anisotropic; in the limit $\delta L\rightarrow 0$, the
transfer matrix $M'\rightarrow I$. This condition will be
implemented by assuming $u'v'=I$. In addition, instead of
modelling the full $p_{\delta L}(\lambda',v')$, we use the
following averages over $M'$:
\begin{equation}
\left<\sum_{a}\lambda'_av'^*_{a\alpha} v'_{a\gamma}
\right>_{\delta L}= \kappa \delta_{\alpha\gamma};
\end{equation}
\begin{equation}
\left<\sum_{ab}\sqrt{\lambda'_a\lambda'_b}v'^*_{a\alpha}v'^*_{a\beta}
v'_{b\delta} v'_{b\gamma}\right>_{\delta L}= \kappa
\delta_{\alpha\gamma}\delta_{\beta\delta}\delta_{\alpha\beta} ,
\end{equation}
where $\kappa=\delta L/l$. The first average is used in
\cite{mpk}, the second one in \cite{chalker}; the latter
incorporates the fact that the thin slice allows backward
scattering without changes in the channel indices, and is highly
anisotropic. Eqs. (10) and (11), with the conditions $u'v'=I$ and
$\kappa \ll 1$, define the model of our building block.

Using the above model and the expansion
$\sqrt{1+\lambda'}=1+\lambda'/2+O(\lambda'^2)$, we average $\delta
\lambda$, Eq. (8), over $M'$:
\begin{equation}
\left<\delta \lambda_{a}\right>_{\delta L}=
\kappa(1+2\lambda_a)+\kappa \sum_{b(\ne a)}
\frac{\lambda_a+\lambda_b+2\lambda_a\lambda_b}{\lambda_a-\lambda_b}
\sum_{\alpha}|v_{a\alpha}|^2|v_{b\alpha}|^2;
\end{equation}
\begin{equation}
\left<\delta \lambda_a\delta\lambda_b\right>_{\delta
L}=\delta_{ab}2\kappa \lambda_a(1+\lambda_a)
\sum_{\alpha}|v_{a\alpha}|^4.
\end{equation}
Similar calculations can be done to obtain the corresponding
averages for $\delta v$.

We should now, in principle, expand Eq. (6) in a Taylor series
around both $\lambda$ and $v$ and evaluate the average  over $M'$.
However, for weak disorder, it is known that the
isotropy approximation is very good, which means $\partial
q_L/\partial v_{ab}\rightarrow 0$. In the strong disorder limit,
the eigenvectors are localized, so that in the the nth row
$v_{na}$ has only one element equal to unity, all other elements
being zero. Since different rows are orthogonal to each other, we
can see from (9) that in this limit $\delta v \rightarrow 0$
because of the restricted sum $m\neq n$. So in both these limits,
the product $(\partial q_L/\partial v_{ab})<\delta v_{ab}>_{\delta
L}\rightarrow 0$. To a good approximation, in these limits, we can
therefore treat $v$ at the macroscopic $L$ as a parameter that
depends on disorder but for a given strength of disorder does not
change any further with $\delta L$. In general, a conductor has a
fraction (depending on the disorder) of its channels closed
\cite{imry}, i.e. the corresponding eigenvectors are localized,
while the others are open, the corresponding eigenvectors being
extended. We expect the isotropy approximation to remain good for
the open channels, and the closed channels to contribute $\delta
v\rightarrow 0$ since they are localized. In other words, the
product $(\partial q_L/\partial v_{ab})<\delta v_{ab}>_{\delta L}$
can be assumed small (compared to the contribution from $(\partial
q_L/\partial \lambda)<\delta \lambda>_{\delta L}$) for all
channels for a wide range of intermediate disorder as well.
Physically, since the eigenvalues depend on the length
exponentially, $\delta \lambda(\lambda,v)$ remains important for
all lengths at any disorder. On the other hand, we expect that at
a macroscopic length $L$, the eigenvectors already evolve to
either metallic (isotropic) or insulating (localized) structures
for a given macroscopic disorder, and any further change due to
$\delta L$ (as opposed to change in disorder) is likely to be
negligible. We will therefore expand the ensemble averaged
marginal probability density $q_L(\lambda,v)$ within the
approximation
\begin{equation}
\left<q_L(\lambda+\delta\lambda(\lambda,v),v+\delta v(\lambda,v))
\right>_{\delta L}\approx
\left<q_L(\lambda+\delta\lambda(\lambda,v),v) \right>_{\delta L}.
\end{equation}
We show below that this approximation retains the dominant
eigenvector correlations (via $\delta\lambda(\lambda,v)$) needed
to reproduce the Lyapunov exponents in arbitrary dimensions as
obtained in \cite{chalker}. The price we pay for not including
$\delta v$ in our calculation is that we will not be able to
evaluate the eigenvector correlations self consistently, but will
have to use them as phenomenological parameters.

Using (14), we expand (6) in a Taylor series about $\lambda$,
\begin{equation}
q_{L+\delta L}(\lambda,v) \approx
q_{L}(\lambda,v)+\sum_a\frac{\partial q_{L} (\lambda,v)}{\partial
\lambda_a} \left< \delta\lambda_a\right>_{\delta L} +\frac{1}{2}
\sum_{ab}\frac{\partial^2 q_{L}(\lambda,v)}
{\partial\lambda_a\partial\lambda_b} \left<
\delta\lambda_a\delta\lambda_b\right>_{\delta L} +\cdots,
\end{equation}
where the dots include terms containing higher order derivatives
of $\lambda$. We choose $\kappa=\delta L/l \ll 1$, which allows us
to truncate the Taylor series at the third term, neglecting terms
of $O(\kappa^2)$ and higher. Using Eqs. (12) and (13), the reduced
distribution $\bar{p}_{L}(\lambda)= \int q_{L}(\lambda,v)d\mu(v)$
can then be written as
$$
\bar{p}_{L+\delta L}(\lambda)\approx \bar{p}_L(\lambda)
+\kappa\sum_a(1+2\lambda_a)\frac{\partial \bar{p}_L(\lambda)}
{\partial\lambda_a}
$$
$$
+\kappa\sum_{b\ne a}
\frac{\lambda_a+\lambda_b+2\lambda_a\lambda_b}{\lambda_a-\lambda_b}
\int\sum_{\alpha}|v_{a\alpha}|^2|v_{b\alpha}|^2
\frac{\partial q_L(\lambda,v)} {\partial\lambda_a} d\mu(v)
$$
\begin{equation}
+ \frac{\kappa}{2} \sum_{a} 2\lambda_a(1+\lambda_a) \int
\sum_{\alpha}|v_{a\alpha}|^4 \frac{\partial^2
q_L(\lambda,v)}{\partial\lambda^2_a} d\mu(v).
\end{equation}

In order to make further progress, we will now make a
{\it`mean-field'} approximation, where the products of four $v$'s
that appear inside the integrals in (16) are replaced by their
mean values which can be taken out of the integrals. This is
equivalent to the assumption that for a given disorder,
fluctuations in such quantities are small compared to their
averages. In the weak disorder regime, each matrix element is of
order $1/\sqrt{N}$, differing mostly in their phases; once the
phases cancel, the fluctuations are negligible for homogeneous
disorder. In the strong disorder regime, each eigenvector has one
element which is unity representing a localized site and the rest
are zero, but different samples will have different localized
sites leading to large sample to sample fluctuations for
individual elements. However, it is expected that the fluctuations
in the sum over the elements of any eigenvector for a given
disorder will remain negligible. Therefore the mean-field
approximation is reasonable in these two limits. As argued before,
we will assume that the assumption remains valid in the
intermediate region of disorder as well, based on the picture of
open and closed channels. Within this approximation, and expanding
the left hand side of Eq. (16) in powers of $\delta L/l$, we get
$$
\frac{\partial \bar{p}_L(\lambda)}{\partial (L/l)} \approx
\sum_a(1+2\lambda_a) \frac{\partial \bar{p}_L(\lambda)}{\partial
\lambda_a}+\sum_{a\ne b}
\frac{\lambda_a+\lambda_b+2\lambda_a\lambda_b}{\lambda_a-\lambda_b}
K_{ab}\frac{\partial \bar{p}_L(\lambda)}{\partial \lambda_a}$$
\begin{equation}
+\sum_{a}\lambda_a(1+\lambda_a) K_{aa}\frac{\partial^2
\bar{p}_L(\lambda)}{\partial \lambda^2_a}.
\end{equation}
Here  we have defined
\begin{equation}
\sum_{\alpha}\left<|v_{a\alpha}|^2|v_{b\alpha}|^2\right>_L\equiv
\sum_{\alpha}\int |v_{a\alpha}|^2|v_{b\alpha}|^2
q_L(\lambda,v)d\mu(\lambda)d\mu(v)\equiv K_{ab}.
\end{equation}

Since $v$ is unitary, $K_{ab}$ satisfies the sum rule $\sum_b
K_{ab}=1$. This allows us to rewrite  Eq. (17), following
[\onlinecite{mu-kl}], as
\begin{equation}
\frac{\partial \bar{p}_L(\lambda)}{\partial (L/l)}
=\frac{1}{\bar{J}}\sum_a
\frac{\partial}{\partial\lambda_a}\left[\lambda_a(1+\lambda_a)K_{aa}
\bar{J}\frac{\partial \bar{p}}{\partial \lambda_a}\right]
\end{equation}
with
\begin{equation}
\bar{J}=\prod_{a<b}|\lambda_a-\lambda_b|^{\gamma_{ab}},\;\;\;
\gamma_{ab}=\frac{2K_{ab}}{K_{aa}}.
\end{equation}
Equation (19), with the definition (20), is our generalization of
the DMPK equation. Note that in the DMPK equation, time reversal
symmetry fixes the exponent in the Jacobian $J$ to be $\beta=1$.
In our case as the eigenvectors $v$
are integrated over, the coupling between the eigenvalues and
eigenvectors adds an effective matrix exponent to the existing symmetry
exponent, resulting in an effective $\bar{J}$ in eq. (19).

Under isotropy condition
$K_{ab}=\frac{1+\delta_{ab}}{N+1}$, we recover the DMPK equation
($\gamma_{ab}=1$). If we choose $K_{ab}=\frac{\mu_1}{N+1}$ and
$K_{aa}=\frac{2\mu_2}{N+1}$, we obtain the generalization of [7],
where an extra condition was needed between $\mu_1$ and $\mu_2$ in
order to satisfy the conservation of probability. In our current
framework, that condition is identical to the sum rule $\sum_b
K_{ab}=1$.

As a check of our model, we evaluate the Liapunov exponents $\nu_a
=\frac{l}{2\delta L} \ln(1+\frac{\delta\lambda_a}{\lambda_a})$ in
the limit $\lambda_1 \gg \lambda_2 \gg \cdots \gg \lambda_N \gg 1$
and compare with [\onlinecite{chalker}]. Expanding in powers of
$\delta\lambda_a/\lambda_a$, averaging over $v$ and using the
results for $<\delta\lambda>$, Eqs. (12-13), we obtain
\begin{equation}
2\nu_a\frac{\delta L}{l} \approx 2\kappa + 2\kappa\sum_{b(\ne
a)}\frac{\lambda_b}{\lambda_a-\lambda_b}
\left<\sum_{\alpha}|v_{a\alpha}|^2|v_{b\alpha}|^2\right>_{L}-
\kappa \left<\sum_{\alpha}|v_{a\alpha}|^4\right>_{L}.
\end{equation}
Separating the sum over $b$ into a term less than $a$ and another
greater than $a$ and using the unitarity of $v$, $\nu_a$ can be
rewritten in the form
\begin{equation}
\nu_a \approx \frac{1}{2}K_{aa}+\sum_{b=a+1}^N K_{ab},
\end{equation}
which coincides with that of [\onlinecite{chalker}]. Note that the
isotropy approximation gives the smallest Liapunov exponent to be
$\nu_N=\frac{1}{N+1}$. While this is the correct form in the
metallic regime ($\lambda_i \ll 1$), the approximation fails to
reproduce the expected behavior in the insulating regime,
$\nu_N\sim O(1)$ independent of $N$. Our approach incorporates the
necessary eigenvector correlations to describe the Liapunov
exponents at weak as well as strong disorder limits. Known
properties of these exponents should provide a guide for
constructing a model of $K$.

It is important to note that while Eq. (19) is of the same form as
the DMPK equation, the presence of the matrix $K$ might not allow
a solution of (19) in the same generic form as that of the DMPK
equation \cite{beenakker}.

In summary, the DMPK equation for the distribution $p_L(\lambda)$
was obtained by adding a thin slice of length $\delta L$ to a
conductor of macroscopic length $L$. The assumption that $p_L$
depends only on $\lambda$ and not on $(u,v)$ limited the equation
to Q1D only. We consider the marginal distribution where the $u$
and $v$ are integrated over. The $u$ integral is done exactly. We
then assume that while the changes in $\lambda$ due to the added
slice depend crucially on the eigenvectors $v$, the eigenvectors
themselves do not change much with length, i.e. they remain either
metallic or insulating as already determined at length $L$. This
implies that our equation is valid only beyond the relaxation
length of the parameters $K_{ab}$. We also assume that the
eigenvector correlations
$\sum_{\alpha}|v_{a\alpha}|^2|v_{b\alpha}|^2$ have sharp
distributions. While these assumptions were made to obtain the
simplest generalization that captures the essentials of
dimensionality dependence and need to be verified independently
(e.g. numerically), we show that our approach keeps the dominant
eigenvector correlations that reproduces the Liapunov exponents in
higher dimensions at both weak and strong disorder. The fact that
the parameters $K_{ab}$ may incorporate proper dimensionality
dependence has already been shown in [\onlinecite{chalker}].
Finally, equation (19) reduces to the DMPK equation as well as to
an earlier generalization [\onlinecite{mu-kl}] in appropriate
limits.

We are grateful to P. Mello for his continued interest and
valuable criticism. We thank Instituto de F\'{\i}sica, UNAM,
M\'exico, where this work originated, for their hospitality. KAM
acknowledges stimulating discussions with J. Chalker and P.
W\"{o}lfle during his stay at U. Karlsruhe where part of the work
was done, supported by the A.v. Humboldt Foundation and the
Max-Planck Society. VAG is grateful for the financial support from
CONACYT, M\'exico, during his stay at UF.

\end{document}